\documentclass[sigconf,screen,authorversion]{acmart}

\usepackage{amsmath,amsfonts}
\usepackage{hyperref}

\newsavebox{\mybox}

\usepackage{amsthm}
\usepackage{mdframed}
\usepackage{centernot}
\usepackage{comment}

\newcommand{\Pp}{\mathcal{P}}
\newcommand{\sPp}{\sem{\Pp}}

\newcommand{\Ss}{\Sigma}

\newcommand{\sem}[1]{ [ \! [ {#1}  ]  \! ]} 

\newcommand\vd[2]{d_{i, p}}
\newcommand{\set}[1]{\left\{ #1 \right\}}

\newcommand{\seq}[1]{\langle #1 \rangle}
\newcommand{\Nat}{\mathbb N}

\newcommand{\R}{\mathbb R}

\newcommand{\Real}{\R}
\newcommand{\Rplus}{\R_{\geq 0}}

\newcommand{\toolname}{\textsc{QFuzz}\xspace}
\newcommand{\diffuzz}{\textsc{DifFuzz}\xspace}
\newcommand{\blazer}{\textsc{Blazer}\xspace}
\newcommand{\themis}{\textsc{Themis}\xspace}
\usepackage{tikz}

\newtheorem{proposition}{Proposition}[section]
\newtheorem{definition}{Definition}[section]

\definecolor{gold}{rgb}{0.99,0.78,0.07}
\usetikzlibrary{arrows,shapes,snakes,automata,backgrounds,positioning,decorations.pathmorphing,
	decorations.markings,calc}

\tikzstyle{dtreenode}=[draw=blue!10!gray,rounded rectangle, minimum size=5mm,fill=blue!10!white]
\tikzstyle{dtreeleaf}=[draw=black!60,minimum width=1cm,minimum height=0.4cm,rectangle,fill=blue!50!white]
\tikzset{every loop/.style={looseness=7}}
\tikzset{
	gluon/.style={decorate,draw=black,
		decoration={coil,amplitude=1pt, segment length=5pt}}
}
\tikzset{
	gluon1/.style={decorate,draw=black,
		decoration={coil,amplitude=3pt, segment length=3pt}}
}
\tikzset{
	gluonew/.style={decorate,draw=black,
		decoration={coil,amplitude=1pt, segment length=2pt}}
}

\tikzset{bicolor/.style args={#1 and #2 and #3}{
		path picture={
			\tikzset{rounded corners=0}
			\fill [#1] (path picture bounding box.south west)
			rectangle
			($(path picture  bounding box.north west)!#3!(path picture bounding
			box.north east)$);
			\fill [#2]
			($(path picture bounding box.south west)!#3!(path picture bounding
			box.south east)$)
			rectangle (path picture bounding box.north east);
}}}

\tikzset{tricolor/.style args={#1 and #2 and #3 and #4 and #5}{
		path picture={
			\tikzset{rounded corners=0}
			\fill [#1] (path picture bounding box.south west)
			rectangle
			($(path picture  bounding box.north west)!#4!(path picture bounding
			box.north east)$);
			\fill [#2]
			($(path picture bounding box.south west)!#4!(path picture bounding
			box.south east)$)
			rectangle
			($(path picture  bounding box.north west)!#5!(path picture bounding
			box.north east)$);
			\fill [#3]
			($(path picture bounding box.south west)!#5!(path picture bounding
			box.south east)$)
			rectangle (path picture bounding box.north east);
}}}

\usepackage{tcolorbox}
\tcbuselibrary{listings,skins}

\lstdefinestyle{mystyle}{
  xleftmargin=0pt,
   basicstyle={\footnotesize\ttfamily},
   aboveskip=3mm,
   belowskip=3mm,
   keywordstyle=\bfseries,
   showstringspaces=false,
  escapechar=?,
  language=Java
}
\definecolor{code_indent}{HTML}{CCCCCC}

\newcommand{\indentrule}{\color{code_indent}\vrule\hspace{2pt}}

\newtcblisting{mylisting}[2][]{
  arc=0pt, outer arc=0pt,
  listing only,
  listing style=mystyle,
  title={\large #2},
  #1
}

 \definecolor{dkgreen}{rgb}{0,0.6,0}
 \definecolor{gray}{rgb}{0.5,0.5,0.5}
 \definecolor{mauve}{rgb}{0.58,0,0.82}

\definecolor{cadmiumgreen}{rgb}{0.0, 0.42, 0.24}
\definecolor{verde}{rgb}{0.25,0.5,0.35}
\definecolor{jpurple}{rgb}{0.5,0,0.35}
\definecolor{darkgreen}{rgb}{0.0, 0.2, 0.13}

\usepackage[ruled,vlined,linesnumbered,lined]{algorithm2e}
\usepackage{algpseudocode}

\usepackage{mathtools,xparse}

\usepackage{tabu}
\usepackage{multirow}

\usepackage{pifont}
\newcommand{\cmark}{\ding{51}}%
\newcommand{\xmark}{\ding{55}}%

\usepackage{enumerate}
\usepackage[shortlabels]{enumitem}

\usepackage{tcolorbox}

\usepackage{pgfplotstable}
\usepackage{pgfplots}

\def\BibTeX{{\rm B\kern-.05em{\sc i\kern-.025em b}\kern-.08em
    T\kern-.1667em\lower.7ex\hbox{E}\kern-.125emX}}

\title{QFuzz: Quantitative Fuzzing for Side Channels}

\setcopyright{acmcopyright}
\acmPrice{15.00}
\acmDOI{10.1145/3460319.3464817}
\acmYear{2021}
\copyrightyear{2021}
\acmSubmissionID{issta21main-p209-p}
\acmISBN{978-1-4503-8459-9/21/07}
\acmConference[ISSTA '21]{Proceedings of the 30th ACM SIGSOFT International Symposium on Software Testing and Analysis}{July 11--17, 2021}{Virtual, Denmark}
\acmBooktitle{Proceedings of the 30th ACM SIGSOFT International Symposium on Software Testing and Analysis (ISSTA '21), July 11--17, 2021, Virtual, Denmark}

\begin{document}

\author{Yannic Noller}
\orcid{0000-0002-9318-8027}
\email{yannic.noller@acm.org}
\affiliation{%
	\institution{National University of Singapore}
	\country{Singapore}
}

\author{Saeid Tizpaz-Niari}
\orcid{0000-0002-1375-3154}
\email{saeid@utep.edu}
\affiliation{%
	\institution{University of Texas at El Paso}
	\country{USA}
}

\begin{abstract}
Side channels pose a significant threat to the confidentiality of software systems.
Such vulnerabilities are challenging to detect and evaluate because they arise from non-functional properties of software such as execution times and require reasoning on multiple execution traces.
Recently, \textit{noninterference} notions have been adapted in static analysis, symbolic execution, and greybox fuzzing techniques.
However, noninterference is a strict notion and may reject security even if the strength of information leaks are weak.
A quantitative notion of security allows for the relaxation of noninterference and tolerates small (unavoidable) leaks.
Despite progress in recent years, the existing quantitative approaches have scalability limitations in practice.

In this work, we present \toolname, a greybox fuzzing technique to quantitatively evaluate the strength of side channels with a focus on \textit{min entropy}.
Min entropy is a measure based on the number of distinguishable observations (partitions) to assess the resulting threat from an attacker who tries to compromise secrets in one try.
We develop a novel greybox fuzzing equipped with two partitioning algorithms that try to maximize the number of distinguishable observations and the cost differences between them.

We evaluate \toolname on a large set of benchmarks from existing work and real-world libraries (with a total of $70$ subjects).
\toolname compares favorably to three state-of-the-art detection techniques.
\toolname provides quantitative information about leaks beyond the capabilities of all three techniques.
Crucially, we compare \toolname to a state-of-the-art quantification tool and find that \toolname significantly outperforms the tool in scalability while maintaining similar precision.
Overall, we find that our approach scales well for real-world applications and provides useful information to evaluate resulting threats.
Additionally, \toolname identifies a zero-day side-channel vulnerability in a security critical Java library that has since been confirmed and fixed by the developers.
\end{abstract}



\maketitle

\section{Introduction}
\label{sec:introduction}

\begin{figure*}
	\centering \includegraphics[width=\linewidth]{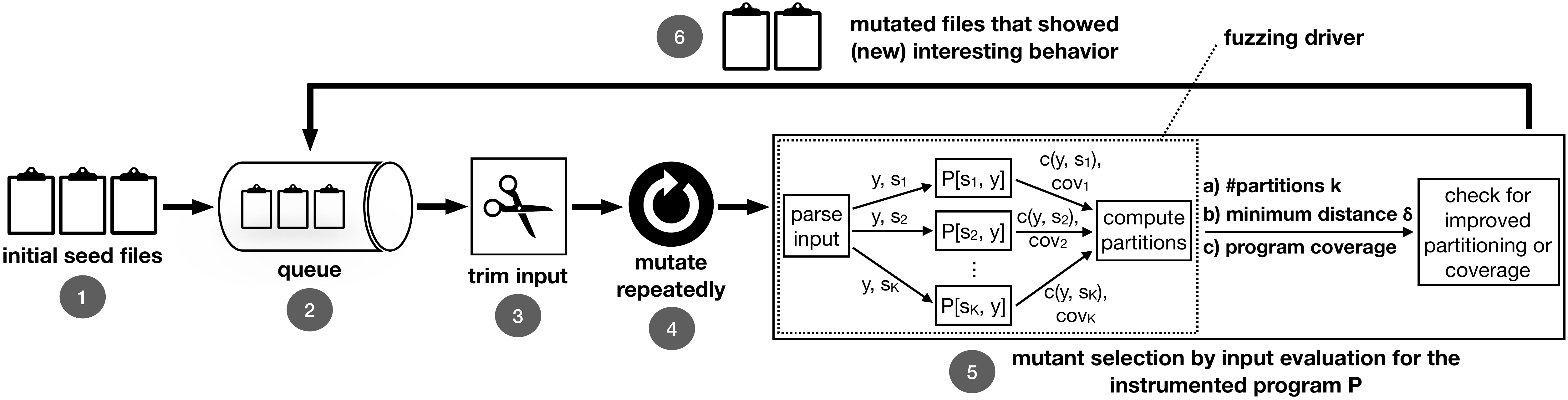}
	\caption{Workflow of \toolname.}
	\label{fig:overview}
\end{figure*}

Side-channel (SC) vulnerabilities allow attackers to compromise secret information by observing runtime behaviors such as response time, cache hit/miss, memory consumption, network packet, and power usage.
Software developers are careful to prevent malicious eavesdroppers from accessing secrets using techniques such as encryption.
However, these techniques often fail to guarantee security in the presence of side channels since they arise from non-functional behaviors and
require simultaneous reasoning over multiple runs.

Side-channel attacks remain a challenging problem even in se\-curity-critical applications.
There are known practical side-channel attacks against the RSA algorithm~\cite{brumley2011remote}, an online health system~\cite{CWWZ10}, the Google's Keyczar Library~\cite{Keyczar}, and the Xbox 360~\cite{xbox}.
In the Xbox 360, timing side channels allowed attackers to reduce the maximum number of trials to compromise a $16$ byte secret from $256^{16}$ to $256*16$ due to vulnerable implementations in byte array comparisons.
Another example is Spectre~\cite{Kocher2018spectre} that challenged the confidentiality of computer devices via side channels.

Recently, techniques have been developed to detect side-channel vulnerabilities in software~\cite{DBLP:conf/icse/nilizadeh,DBLP:conf/ccs/ChenFD17,antonopoulos2017decomposition}.
These works have shown notable success in finding critical vulnerabilities in libraries such as Eclipse Jetty, Apache Ftp, and OpenJDK Crypto.
Despite these discoveries, the detection techniques often rely on the \textit{noninterference} notion of security, which leads to binary answers with limited extra information such as the maximum timing difference observed to reject the security~\cite{DBLP:conf/icse/nilizadeh}.
A quantitative notion is crucial in the security evaluation of real-world applications.
The notion allows developers to evaluate the resulting threat of side channels precisely and even relax noninterference to tolerate small leaks, which might be necessary in practice.
%
Previous work has also used quantitative methods to evaluate the strength of information leaks~\cite{smith2009foundations,backes2009automatic,pasareanu2016multi},
however, they are limited to small programs and do not scale well for large applications and input spaces.

Prevalent quantification measures such as Shannon entropy~\cite{KB07,backes2009automatic}
show the expected amount of leaks over an unbounded number of trials and fail
to evaluate the resulting threats for a more practical setting where an attacker can try one ideal guess.
In this paper, we focus on the immediate threats that allow an attacker to guess the secret correctly in one try.
In this threat model, Smith~\cite{smith2009foundations} showed that
the number of distinguishable observations (i.e., partitions)
precisely quantifies the strength of information leaks.
The corresponding quantitative notion is known as \textit{min entropy}.

We introduce a practical variant of min entropy that
can tolerate $\epsilon$ cost differences in introducing a new partition.
Then, we adapt a greybox evolutionary algorithm to approximate the number of
partitions with lower-bound guarantees.
To the best of our knowledge, this is the first work to adapt greybox evolutionary fuzzing to characterize quantitative measures of information leaks that scale well for large applications, handle dynamic features, and provide lower-bound guarantees.
In particular, we extend \diffuzz~\cite{DBLP:conf/icse/nilizadeh}, a greybox fuzzer for side-channel detection, with partitioning algorithms to quantify the amount of leaks.
In addition to code coverage measures, our algorithm guides the search to find as many secret values along with single public value such that the following criteria are maximized:
(1) the number of partitions and
(2) the cost differences between partitions.

We propose two partitioning algorithms to find distinguishable classes and the cost distance between them.
The first algorithm takes a dynamic programming approach and maximizes both criteria, but it may not provide bounds on the estimation of information leaks.
The second partitioning algorithm is greedy, which has cheaper computations and provides a lower-bound guarantee, but it may not maximize the cost differences between partitions.

We implement our approach in a tool named \toolname and apply \toolname on $70$ subjects including real-world \textsc{Java} libraries.
On the set of benchmarks, we first compare the two partitioning algorithms.
Then, we compare our approach against state-of-the-art detection tools.
The result shows that \toolname has better scalability and detection when compared to \themis~\cite{DBLP:conf/ccs/ChenFD17} and \blazer~\cite{antonopoulos2017decomposition}, while it is similar to \diffuzz~\cite{DBLP:conf/icse/nilizadeh}.
\toolname also provides quantitative information that is useful to evaluate threats, beyond the capabilities of \themis, \blazer, and \diffuzz.
Finally, we compare our approach to the quantification technique \textsc{MaxLeak}~\cite{pasareanu2016multi}.
\toolname outperforms \textsc{MaxLeak} in scalability while providing comparable precision.

On a set of real-world benchmarks, we show the scalability and usefulness
of our approach. \toolname discovers a zero-day side-channel vulnerability
in Apache WSS4J, a library for the implementation of Web Services Security.
This vulnerability has been confirmed and fixed by developers~\cite{WSS4J}.

\begin{tcolorbox}[boxrule=1pt,left=1pt,right=1pt,top=1pt,bottom=1pt]
\textbf{Our main contributions are:}
\begin{itemize}[leftmargin=1.5em,rightmargin=0em]
	\item a novel greybox fuzzing approach to \textit{detect} and \textit{quantify} side-channel vulnerabilities by characterizing the min entropy,
	\item the publicly available implementation of \toolname, and
	\item the evaluation of \toolname by showing its scalability and usefulness when compared to state-of-the-art techniques as well as in real-world applications, including the discovery of a previously unknown vulnerability
	in a security-critical library.
\end{itemize}
\end{tcolorbox}

\section{Overview}
\label{sec:overview}
\begin{figure*}[t!]
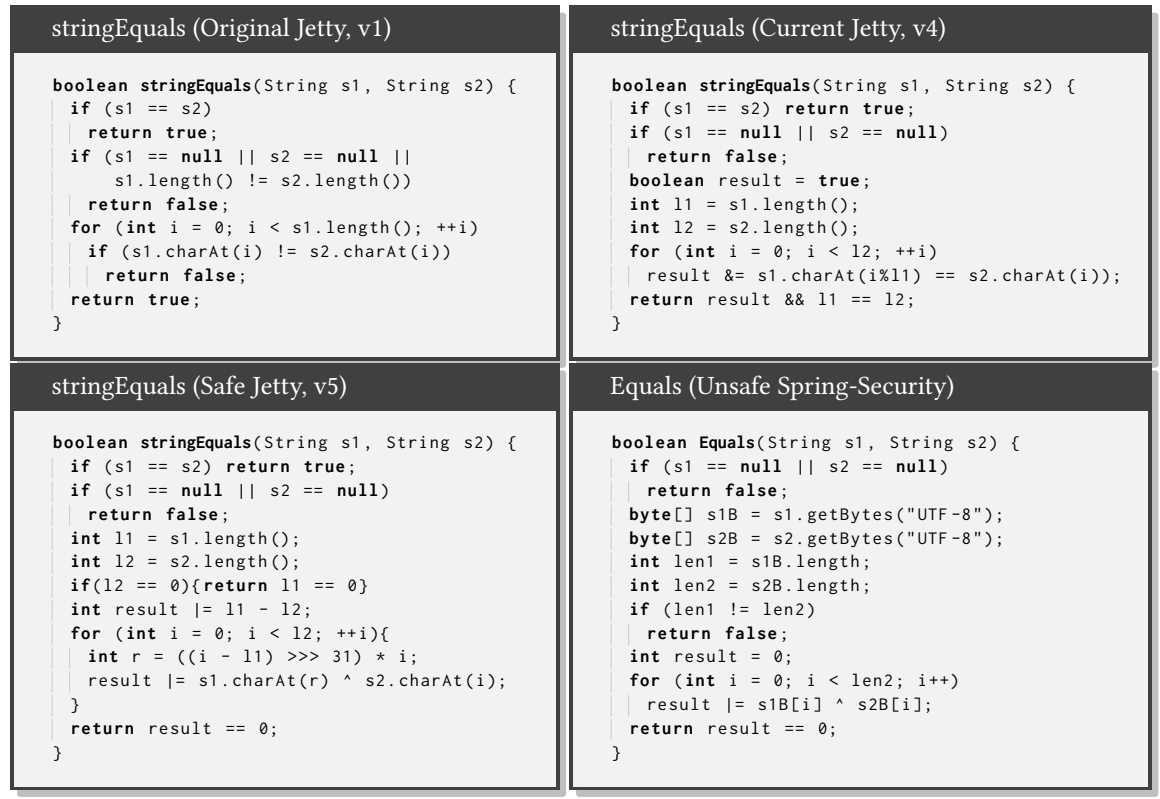

\centering
\begin{lrbox}{\mybox}%
\begin{scriptsize}
\begin{mylisting}[hbox,enhanced,drop shadow]{stringEquals (Original Jetty, v1)}
boolean ?{\textbf{stringEquals}}?(String s1, String s2) {         .
?\indentrule? if (s1 == s2)
?\indentrule? ?\indentrule? return true;
?\indentrule? if (s1 == null || s2 == null ||
?\indentrule?     s1.length() != s2.length())
?\indentrule? ?\indentrule? return false;
?\indentrule? for (int i = 0; i < s1.length(); ++i)
?\indentrule? ?\indentrule? if (s1.charAt(i) != s2.charAt(i))
?\indentrule? ?\indentrule? ?\indentrule? return false;
?\indentrule? return true;
}
\end{mylisting}
\end{scriptsize}
\end{lrbox}%
\scalebox{1.0}{\usebox{\mybox}}
\begin{lrbox}{\mybox}%
\begin{scriptsize}
\begin{mylisting}[hbox,enhanced,drop shadow]{stringEquals (Current Jetty, v4)}
boolean ?{\textbf{stringEquals}}?(String s1, String s2) {
?\indentrule? if (s1 == s2) return true;
?\indentrule? if (s1 == null || s2 == null)
?\indentrule? ?\indentrule? return false;
?\indentrule? boolean result = true;
?\indentrule? int l1 = s1.length();
?\indentrule? int l2 = s2.length();
?\indentrule? for (int i = 0; i < l2; ++i)
?\indentrule? ?\indentrule? result &= s1.charAt(i%l1) == s2.charAt(i));
?\indentrule? return result && l1 == l2;
}
\end{mylisting}
\end{scriptsize}
\end{lrbox}%
\scalebox{1.0}{\usebox{\mybox}}
\begin{lrbox}{\mybox}%
\begin{scriptsize}
\begin{mylisting}[hbox,enhanced,drop shadow]{stringEquals (Safe Jetty, v5)}
boolean ?{\textbf{stringEquals}}?(String s1, String s2) {         .
?\indentrule? if (s1 == s2) return true;
?\indentrule? if (s1 == null || s2 == null)
?\indentrule? ?\indentrule? return false;
?\indentrule? int l1 = s1.length();
?\indentrule? int l2 = s2.length();
?\indentrule? if(l2 == 0){return l1 == 0}
?\indentrule? int result |= l1 - l2;
?\indentrule? for (int i = 0; i < l2; ++i){
?\indentrule? ?\indentrule? int r = ((i - l1) >>> 31) * i;
?\indentrule? ?\indentrule? result |= s1.charAt(r) ^ s2.charAt(i);
?\indentrule? }
?\indentrule? return result == 0;
}
\end{mylisting}
\end{scriptsize}
\end{lrbox}%
\scalebox{1.0}{\usebox{\mybox}}
\begin{lrbox}{\mybox}%
\begin{scriptsize}
\begin{mylisting}[hbox,enhanced,drop shadow]{Equals (Unsafe Spring-Security)}
boolean ?{\textbf{Equals}}?(String s1, String s2) {             ?{\color{white}.}?
?\indentrule? if (s1 == null || s2 == null)
?\indentrule? ?\indentrule? return false;
?\indentrule? byte[] s1B = s1.getBytes("UTF-8");
?\indentrule? byte[] s2B = s2.getBytes("UTF-8");
?\indentrule? int len1 = s1B.length;
?\indentrule? int len2 = s2B.length;
?\indentrule? if (len1 != len2)
?\indentrule? ?\indentrule? return false;
?\indentrule? int result = 0;
?\indentrule? for (int i = 0; i < len2; i++)
?\indentrule? ?\indentrule? result |= s1B[i] ^ s2B[i];
?\indentrule? return result == 0;
}
\end{mylisting}
\end{scriptsize}
\end{lrbox}%
\scalebox{1.0}{\usebox{\mybox}}

\caption{String equality in Eclipse Jetty ({\tt s1} secret, {\tt s2} public).
Top-Left: The code snippet is the original implementation for the secret comparison that contains a strong side channel.
Top-Right: The code is the current version that has been developed to fix the side channel, but still leaks some information.
Bottom-Left: The code snippet is a proposed safe implementation.
Bottom-Right: String equality in Spring-Security that leak whether the length of strings is matching.}
\label{fig:jetty-streql}
\end{figure*}

\subsubsection*{\toolname in a Nutshell}
%
\toolname searches for a single public input and $K$ secret inputs that maximize the number of partitions, as well as the distance between these partitions.
Intuitively, a pair of secret inputs is distinguishable and belongs to different partitions if their cost differences are greater than a tolerance parameter $\epsilon$.
As illustrated in Figure~\ref{fig:overview}, \toolname uses an evolutionary greybox fuzzing approach to evolve the public and secret values.
It starts with some initial seed files and applies random mutations like random bit flips and crossovers.
To assess the quality of the evolved inputs, the fuzzing driver parses the inputs into a public value $y$ and $K$ secret values ($s_1, s_2, ..., s_K$), and executes the instrumented program with the corresponding input pairs.
The program's instrumentation keeps track of the resource usage (e.g., number of executed instructions, or memory consumption), and provides the actual observation, denoted as the function $c$ in Figure~\ref{fig:overview}.
Afterward, the observations are processed and the corresponding partitions are computed.
Finally, the fitness function decides based on the number of partitions in the side-channel observations, the minimum distance $\delta$ between these partitions, and the overall coverage whether it is interesting to keep the input.

\subsubsection*{Example}
We use the password matching implementations from Eclipse Jetty as well as Spring-Security
to illustrate our approach.
Figure \ref{fig:jetty-streql} shows three different variants of Jetty and
an unsafe variant of Spring-Security.
The first variant was vulnerable to timing side channels since it used \textsc{Java} internal string equality to compare a (secret) password against a given (public) guess as shown in Figure~\ref{fig:jetty-streql} (top-left).
The developers made multiple fixes and finally committed the final fix~\cite{jetty-3} (the current implementation) as shown in Figure~\ref{fig:jetty-streql} (top-right).
The implementation in Figure~\ref{fig:jetty-streql} (bottom-left) is another candidate for password matching, taken from OpenJDK library~\cite{openjdk-fix}. Finally, we consider an unsafe variant of password matching from Spring-Security as shown in Figure~\ref{fig:jetty-streql} (bottom-right).
We study the feasibility of side channels in these four implementations
and apply \toolname to estimate the amount of information leaks using the number of partitions.

\subsubsection*{Example Parameters}
We consider $K$ = $100$ and $\epsilon=1$ as default configuration parameters.
We set the length of the secret and the public guess to be the same and fixed to $16$ characters.
We run \toolname $30$ times on each variant, where each run
is for $30$ minutes. We report the maximum number of partitions ($k$)
and the cost differences in bytecode between two closest partitions ($\delta$).
The detailed results for can be found in Table~\ref{table:benchmark-cluster} and Table~\ref{table:benchmark-themis}.

\subsubsection*{First Variant of Jetty}
Figure~\ref{fig:jetty-streql} (top-left) shows the first variant, for which \toolname discovers $17$ classes of observations ($k=17$).
Each partition is at least $3$ bytecodes far from any other partition ($\delta=3$).
Since we fix the length, the number of partitions reflect side-channel observations related to the content of secret inputs.
We find that each partition shows the number of characters in the prefix of
secrets that match with the guess.
Since there are $16$ characters, there can be $17$ partitions ranging from no prefix match to all $16$ characters match.
This implementation is known to be vulnerable to adaptive side channels where
an attacker can use the cost observations to compromise a prefix of a
secret password in each step of the attack.
The outcome of \toolname indirecly indicates the feasibility of adaptive attacks,
while they are not the main focus.

\subsubsection*{Second Variant of Jetty (current implementation)}
We consider the current implementation in Jetty as shown in
Figure~\ref{fig:jetty-streql} (top-right).
In this case, \toolname detects $9$ partitions where each partition is at least $1$ bytecode far from any others.
This analysis shows that the fix improved the security and reduced the strength of leaks as compared to the first variant.
Since \toolname found multiple partitions, however, we conclude that this variant is not completely safe.
To understand the issue, we analyze the corresponding instructions generated by \textsc{Java} Virtual Machine (JVM).
The analysis shows the equal operator (``=='') in the loop body is optimized
by JVM and translated to a conditional jump instruction (\texttt{if\_icmpne}) if the comparison is not successful and an unconditional jump instruction otherwise.
This translation introduces an imbalance comparison where the unconditional jump includes a single extra bytecode instruction as compared to the conditional jump.
With $16$ characters, the bytecode differences, range from $0$ to $16$, are partitioned into $9$ classes with $\epsilon=1$.

\subsubsection*{Third Variant of Jetty (OpenJDK~\cite{openjdk-fix})}
We take a password matching algorithm from OpenJDK~\cite{openjdk-fix} that
explicitly uses ``xor'' operation instead of ``=='' as depicted in
Figure~\ref{fig:jetty-streql} (bottom-left).
We apply \toolname to this implementation and obtain only $1$ partition.
Therefore, we deem this variant a completely safe implementation for the password matching.
Looking into the instruction bytecodes, we did not find any conditional or unconditional jump in the loop body.
The JVM directly uses the xor operation for the comparison.

\subsubsection*{Unsafe Variant of Spring-Security}
We study an unsafe string matching algorithm from \themis~\cite{DBLP:conf/ccs/ChenFD17},
especially to compare our approach with the state-of-the-art. As shown in
Figure~\ref{fig:jetty-streql} (bottom-right)
the length of secret and public strings are compared after encoding with the
\texttt{String.getBytes("UTF-8")} function.
We apply \toolname to this implementation, and it reported $2$ partitions
with $\delta=149$. With a closer look into the identified partitions,
the secret values in one partition contain a special character,
which is then mapped to two byte values. This indicates that
there is an early return of length mismatch over bytecode encodings even if the string
lengths of secret and public inputs are the same.

\subsubsection*{Comparison to Related Work}
The state-of-the-art \textit{detection} techniques such as \diffuzz~\cite{DBLP:conf/icse/nilizadeh}
rely on the noninterference that deems applications safe or unsafe with
extra information such as the maximum cost difference between two secret inputs,
which are not as useful as the number of partitions.
We use the example of Spring-Security, Figure~\ref{fig:jetty-streql} (bottom-right),
to show differences concretely.
\toolname identifies $2$ partitions with $\delta=149$.
Similarly, \diffuzz classifies the subject as unsafe with the cost differences of $149$ bytecodes.
At first glance, the results look similar.
But, there are subtle differences.
\diffuzz implies a strong side channel by showing that the maximum differences
between two secret values are $149$ bytecodes.
\toolname indicates that there are
only two partitions that are $149$ bytcodes far from each other, and there is no
other observation between them with cost differences more than one bytecode ($\epsilon=1$).
Thus, \toolname implies that the strength of leak is weak and
is unlikely to compromise whole secret. In fact,
the side channel only leaks whether the secret string contains a special
character or not.
%

Additionally, the state-of-the-art \textit{quantitative} techniques for side channels like
\textsc{MaxLeak}~\cite{pasareanu2016multi} are often developed based on
static analysis or symbolic execution, which may not scale for practical examples as shown in Section~\ref{subsec:results-rsa}.

\subsubsection*{Summary}
Our analysis provides an under-approximation of the (true) number of partitions via
side-channel observations.
As shown in this example, \toolname provides quantitative information that can be useful to evaluate security among multiple variants, understand the strength of leaks,
and identify safer implementations that meet the security requirements.

\section{Problem Definition}
\label{sec:definition}
First, we define the resource usage model of programs:
\begin{definition}[Cost Model]
  The cost model of a deterministic program $\Pp$ is a tuple
  $\sPp = (X, Y, \Ss, c)$ where
  $X = \set{x_1, \ldots, x_n}$ is the set of secret inputs,
  $Y = \set{y_1, \ldots, y_m}$ is the set of public inputs,
  $\Ss \subseteq \Real^n$ is a finite set of secrets, and
  $c: \Real^n \times \Real^m \to \Rplus$ is the cost function of the program over the secret and public inputs.
\end{definition}

We assume that the abstraction of execution times as the number of executed \textsc{Java} bytecodes is precise and corresponds to the actual timing observations.
Therefore, we overwrite the cost function to be
$c: \Real^n \times \Real^m \to \Nat_{>0}$.
Our goal is to characterize the information leaks in a given program
with the quantitative measures. Thus, we consider the
finite set of secret inputs.

Smith~\cite{smith2009foundations} noted that prevalent
quantitative metrics such as Shannon entropy
characterize the expected resulting threats
over unbounded number of trials. Instead, Smith
proposed \textit{min} entropy that reflects the resulting
threat from an attacker who can try one best guess.
To see the differences,
let us consider the timing models of two programs
$T_A = H~|~31$ and $T_B = H~\&~L$ where $H$ and $L$ are $6$ bits (uniform)
secret and public inputs, respectively. Program $A$ leaks the most significant bit,
and program $B$ leaks zero to all bits depending on $L$.
\textit{Shannon} entropy~\cite{KB07}
quantifies the leaks to be $1$ bit and $3$ bits for the first and
second variants, respectively. On the other hand,
min entropy~\cite{smith2009foundations} uses the maximum partitions and quantifies the
leaks to be $\log_2{2}=1$ bit
and $\log_2{64}=6$ bits for the first and second variants, respectively!
We note that $64$ partitions can be observed with the
best guess of $L=63$ in the program $B$.

\subsubsection*{Threat Model}
We consider a chosen-message threat setting~\cite{kopf2009provably} where
an attacker can pick an ideal public input to compromise the secret value or some
properties of it in one try. In the offline mode, the attacker, who
has access to the source code, samples execution times
using the ideal public input and partitions secret values into
different classes of observations.
During the online phase of the attack, the attacker queries the target
application with the public input and tries to compromise the
fixed secret by mapping the observed time to a partition of secret values.
We assume that the precision in the attacker's observation can be
set with a tolerance parameter $\epsilon$ such that any cost differences
more than $\epsilon$ is distinguishable.

\begin{proposition}[Smith~\cite{smith2009foundations}]
\label{prop:min-entropy}
Let $\Ss_{Y=y} = \seq{S_1, S_2, \ldots, S_k}$ be
the quotient space of $\;\Ss$ characterized by the cost observations under
the public input $y$ such that any pair of secret values are
in the partition $i$ ($s,s' \in S_i$ for any $1 \leq i \leq k$), if and only
if $c(s,y) = c(s',y)$.
Let $Y=y^*$ be a single public input that gives the maximum
number of classes in the cost observations $k^*$, i.e.,
$|\Ss_{Y=y^*}| \geq |\Ss_{Y=y}|$ for any $y \in Y$.
Assuming that the program $\Pp$ is deterministic and the distribution over
secret input $\Ss$ is uniform, then amounts of information leaked
according to min-entropy measure can be characterized with $log_2 k^*$.
\end{proposition}

Proposition~\ref{prop:min-entropy} characterizes min entropy with the exact
cost equality, i.e., $\epsilon = 0$. However, the exact equality is a
strict security policy, and it is prevalent to
account for uncertainties in observations with
$\epsilon > 0$~\cite{DBLP:conf/ccs/ChenFD17,DBLP:conf/icse/nilizadeh,FuncSideChan20}.
Since a positive value of $\epsilon$ does
not characterize the equivalence class over $\Ss$, we adapt
the proposition~\ref{prop:min-entropy} with a relation that is
reflexive and symmetric, but not transitive. The relation is known
as tolerance relation~\cite{10.1016,10.1145}.

\begin{proposition}[Maximal set of similar secrets~\cite{10.1145}]
\label{prop:tolerance-relation}
Given any tolerance $\epsilon \geq 0$,
a quotient space of $\Ss_{Y=y} = \seq{S_1, S_2, \ldots, S_k}$
characterizes the maximal tolerance classes if and only if
(1) $s,s' \in S_i$ (for $1 \leq i \leq k$),
if and only if $|c(s,y) - c(s',y)| \leq \epsilon$ and
(2) $\forall s \not \in S_i$ (for $1 \leq i \leq k$),
$\exists s' \in S_i$, $|c(s,y) - c(s',y)| > \epsilon$.
\end{proposition}

It is straightforward to prove that the number of maximal tolerance classes is equal to the number of distinguishable observations for an attacker with the tolerance parameter $\epsilon \geq 0$.
In addition to the number of observations,
the cost differences between distinguishable observations are critical to
compromise secrets in one trial.
Given two public inputs $y_1,y_2$ that characterize the same
number of distinguishable observations $k$, we measure the distance between
observations with
\begin{center}
$\xi(\Ss,y) = \sum_{i=1}^{k}\min_{j=i+1}^{k}\set{|c(s,y) - c(s',y)|: s \in S_i \land s' \in S_j}$
\end{center}
and pick $y_1$ over $y_2$ if and only if $\xi(\Ss_1,y_1) > \xi(\Ss_2,y_2)$.



\begin{definition}[Problem Statement]
    \label{def-problem}
    Given a deterministic program $\Pp$ with secret inputs
    $X \in \Real^n$ and public inputs
    $Y \in \Real^m$, a cost function $c \in \Real^n \times \Real^m
    \to \Nat_{>0}$, and a tolerance $\epsilon \in \Nat$,
    the key computational problem is to find a finite set
    of secret values $\Ss$ and a public input $y^*$
    that characterizes the maximum number of classes
    in the observation with the highest distances,
    i.e., $|\Ss_{Y=y^*}| > |\Ss_{Y=y}|$ or $|\Ss_{Y=y^*}| = |\Ss_{Y=y}| \land
    \xi(\Ss,y^*) \geq \xi(\Ss,y)$ for any $y \in Y$, and
    compute min entropy as $log_2 |\Ss_{Y=y^*}|$.
\end{definition}

The problem of finding inputs to characterize the maximum
number of distinguishable observations with the highest distance is hard. First,
the input generation part requires an exhaustive search in the exponential
set of subsets of input space, and hence is clearly intractable.
Given a set of secret inputs $\Ss$, and public input $y$,
finding the number of distinguishable observations with a high distance is NP-complete.
The proof can establish from Graph $K$-Color\-ability problem that is known to
be NP-Complete for any $K \geq 3$.

\section{quantifying leaks with fuzzing}
\label{sec:approach}
Since the problem~\ref{def-problem} is hard,
we propose a dynamic approach to approximate the min entropy with
lower-bound guarantees.
In particular, we develop an evolutionary fuzzing algorithm with
constrained partitioning.


\begin{algorithm}[t!]
{
	\DontPrintSemicolon
	\KwIn{Program $\Pp$,
	initial seed $IS$,
	partitioning algorithm $Part$,
	tolerance parameter $\epsilon$,
	timeout $T$,
	max. num. partitions $K$.
	}
	\KwOut{num. partitions $k$, distance $\delta$}

	$k$, $\delta$, $population$ $\gets$ $1$, $0$, $IS$

	$inputs$ $\gets$ \texttt{mutation\_pick}($population$)

	$y, s_1, \ldots, s_K$ $\gets$ \texttt{parse}($inputs$, $constraints$)

	($cost_1$,$p_1$), \ldots, ($cost_K,p_K$) $\gets$ $\Pp$($s_1$, $y$), $\ldots$,
	$\Pp$($s_K$, $y$)

	$k', \delta'$ $\gets$ \texttt{Part$_{\epsilon}$}($cost_1$, $\ldots$, $cost_K$)

	\If{$k' > k$ \textbf{or} ($k = k'$ \textbf{and} $\delta' > \delta$)}{
		\texttt{Add} ($y, s_1, \ldots, s_K$) to population

		$k$, $\delta$ $\gets$ $k'$, $\delta'$
	}
	\ElseIf{any path $p_i$ characterizes a new path}
	{
		\texttt{Add} ($y,s_i$) to population
	}

	\If{$t \leq T$}{
		Go to $2$
	}
	\Else{
		\Return $k$, $\delta$.
	}
	\caption{\toolname: Evolutionary fuzzing for quantifying information leakage.}
	\label{alg:qfuzz-overall}
}
\end{algorithm}

\subsubsection*{General Algorithm}
Algorithm~\ref{alg:qfuzz-overall} shows the high-level steps in our quantitative fuzzing approach.
A high-level illustration is shown in Figure \ref{fig:overview}.
It starts with picking an input from the population and performs some mutations.
For this step, we build on top of existing fuzzing approaches like AFL \cite{AFL}, which includes, e.g., random bit flips, random deletions/insertions, and crossovers.
After parsing the inputs, \toolname runs the program on the set of secret and public inputs and characterizes the paths and costs.
Then, it applies the partitioning algorithm with the given tolerance parameter and returns the number of partitions and the distance between them.
The fuzzer will keep an input if the values characterize more distinguishable observations or a higher distance between the partitions.
Similar to prevalent evolutionary fuzzing, inputs with new path coverage are still added to the population.


\subsubsection*{Objective Function}
The fuzzing objective is to find a public input and a set of secret inputs such that the number of distinguishable observations over the set of secret values is maximized.
Let $\epsilon \in \Nat_{\geq 0}$ be the tolerance parameter to distinguish secret values and $K$ be an upper-bound on the number of distinguishable observations.
The quantitative fuzzing aims to maximize the following objective and returns corresponding set of secrets and a public input:
\begin{center}
$\max_{s_1,\ldots,s_K,y} |\texttt{Part}_{\epsilon}\big(c(s_1,y),
\ldots,c(s_K,y)\big)|~~+ \big(1 - \exp(-0.1*\delta)\big)$
\end{center}
where the partitioning function \texttt{Part}$_{\epsilon}$ returns the distinguishable classes of observations (partitions) from the generalized set of $K$ secret values and a public input $y$ under the tolerance parameter $\epsilon$. The distance parameter $\delta$ is measured by the \texttt{Part}$_{\epsilon}$ function.
The term $1 - \exp(-0.1*\delta)$ is always between $0$ and $1$, and prefers a partitioning
with the maximum distance $\delta$ over any other partitioning with the same number of classes.
Thus, the objective function of our fuzzer guides the search for finding interesting inputs with two criteria:
(1) the number of discovered partitions is maximized and
(2) the cost differences (distance) between partitions are maximized.

We combine our custom objective function with the code coverage criteria during
fuzzing.
In this way, our approach can circumvent local minima in the search space: if we cannot identify more partitions, our search will still attempt to increase program coverage and continue to explore other parts of the program.

The partitioning algorithm \texttt{Part}$_{\epsilon}$ is a key component in our fuzzing approach.
We propose two algorithms for efficient partitioning, which both have their merits: \textit{KDynamic} and \textit{Greedy}.
The \textit{KDynamic} algorithm (see Section \ref{sec:kdynamic}) adapts a dynamic programming approach to find classes that have large distances.
This algorithm, however, is expensive and may over-approximate the number of distinguishable classes.
The \textit{Greedy} algorithm (see Section \ref{sec:greedy}) performs a greedy selection of partitions.
This algorithm is fast and finds the exact number of partitions, but it may not find a partitioning with the maximum distance between classes.

%

\subsection{Partitioning with Dynamic Programming}
\label{sec:kdynamic}
Our \textit{KDynamic} algorithm is inspired by ideas from computing optimal histograms~\cite{10.5555/645924.671191}.
Let $T = \set{c_1,\ldots,c_K}$ be the set of positive integers (costs) that is indexed in ascending order, i.e., $c_i < c_j$ whenever $i < j$.
Let $T_r = \set{c_1,\ldots,c_r} \subseteq C$ be the set of costs up to the index $r \leq K$.
An $k-$partition of $T$ is $B_k = \set{b_1,\ldots,b_k}$ where each
class $i$ is $\set{c \in T | b_{i-1} < c \leq b_i}$.
We define the cost of partitioning to be the sum of differences
between the smallest and largest costs in each class. Formally,
$cost(B) = \sum_{i=1}^{k} |\max \set{c | c \in b_i} - \min \set{c | c \in b_i}|.$

It follows from~\cite{10.5555/645924.671191,kopf2009provably} that every
optimal partitioning of $T_r$ contains an optimal
partitioning for some $T_q$ with $q < r$. With this optimal substructure
property, we can now define the dynamic programming to construct optimal
partitioning. Let $a(r,i)$ be the cost of partitioning for $T_r$ and $i$ classes
defined as the following:
\begin{equation*}
\begin{split}
a(r,1) = |c_r - c_1|~~~~~~~~~~~~~~~~~~~~~~~~~~\\
a(r,i) = \min_{1 \leq j < r} \big\{a(j,i-1) + |c_r - c_{j+1}|\big\}
\end{split}
\end{equation*}
For a given $k$, the worst-case running time for this algorithm
is $O(k.K^2)$. We invoke this algorithm for each $k=1,\ldots,K$
in this order and find the smallest $k$ to satisfy
the constraint $\forall b_i \in B_k, \forall s,s' \in b_i$,
$|c(s,y)-c(s',y)| \leq \epsilon$,
with the distance between partitions is:
\begin{center}
$\sum_{i=1}^{k-1}$ $|B_i.max{-}B_{i+1}.min|$ for $k{>}1$.
\end{center}

\subsection{Partitioning with Greedy Approach}
\label{sec:greedy}
Our \textit{Greedy} partitioning is shown in Algorithm~\ref{alg:greedy-clustering}.
After preprocessing the set of cost observations, \textit{Greedy} picks the observation with the lowest cost and puts all other observations (in the sorted order) that are $\epsilon$-close to the lowest observation in the first class.
Next, \textit{Greedy} picks the lowest cost observation that is not covered in the first class and puts all $\epsilon$-close observations in the second class.
This procedure continues until all costs are covered and assigned to a class.
The complexity of algorithm is equal to the complexity of the sorting algorithm that is $O(K.\log(K))$.

\section{Evaluation}
\label{sec:experiment}
We evaluate \toolname on a large set of benchmarks and focus thereby on the following three research questions:
\begin{enumerate}[start=1,label={\bfseries RQ\arabic*},leftmargin=3em]

\item Which partitioning algorithm (\textit{KDynamic} or \textit{Greedy}) performs better in terms of correct number of partitions and time for partition computation?

\item How does \toolname compare with state-of-the-art SC detection techniques like \blazer~\cite{antonopoulos2017decomposition}, \themis~\cite{DBLP:conf/ccs/ChenFD17}, and \diffuzz~\cite{DBLP:conf/icse/nilizadeh}?

\item Can \toolname be used for the quantification of SC vulnerabilities in real-world \textsc{Java} applications and how does it compare with \textsc{MaxLeak} \cite{pasareanu2016multi}?

\end{enumerate}
All subjects, experimental results, and our tool are available on our GitHub repository:
\url{https://github.com/yannicnoller/qfuzz}

\begin{algorithm}[t!]
{
	\DontPrintSemicolon
	\KwIn{Cost $T=\set{c_1,\ldots,c_K}$, threshold $\epsilon$}
	\KwOut{num. partitions $k$, distance $\delta$}

	$remove\_duplicates\_sort(T)$

	$B, b, \delta$ $\gets$ $\set{}$, $\emptyset$, 0

	\For{each $c$ $\in$ $T$}
	{
		\If{$b=\emptyset$ \textbf{or} ($|c - b.min| \leq \epsilon$)}{
				$b$.add($c$)
		}
		\Else{
			$B$.add($b$)

			$b$ $\gets$ $\set{c}$
		}
	}
	$B$.add($b$)

	\If{$|B| \geq 2$}{
		$\delta$ $\gets$ $\sum_{i=1}^{|B|-1}$ $|B_i.max - B_{i+1}.min|$
	}

  \Return $|B|$,$\delta$

	\caption{Partitioning with a greedy approach.}
	\label{alg:greedy-clustering}
}
\end{algorithm}

\subsection{Subjects}
\label{sec:subjects}
For RQ1, we use a micro-benchmark and real-world \textsc{Java} programs.
We compare the partitioning algorithms on the variants of \textit{Eclipse Jetty} presented in Section~\ref{sec:overview} and a set of \textit{Leak Set} programs that are leaking the number of set bits.
In \textit{Eclipse Jetty}, the secret inputs are fixed to have a length of $16$ characters.
In \textit{Leak Set}, the size of secrets are from $12$ to $28$ bits for \textit{Leak Set 1} to \textit{Leak Set 5}, respectively.
Additionally, we apply \toolname on \textit{Apache WSS4J}, for which we newly reported a vulnerability that has been fixed by the developers
\footnote{\url{https://issues.apache.org/jira/browse/WSS-677}}.
For RQ2, we apply \toolname on the benchmarks of \blazer, \themis, and \diffuzz.
They include micro-benchmarks and real-world \textsc{Java} programs.
Finally, for RQ3, we apply \toolname on the RSA subjects taken from Pasareanu et al.~\cite{pasareanu2016multi}.
They represent an unsafe implementation of modular exponentiation, which can leak information about the secret exponent via a timing side channel.
We compare our approach to \textsc{MaxLeak}~\cite{pasareanu2016multi} that also quantifies leaks with \textit{min} entropy using symbolic executions and model counting.
%
%

\subsection{Technical Details}
Our tool \toolname, similar to~\diffuzz~\cite{DBLP:conf/icse/nilizadeh},
is implemented on top of \textsc{AFL}~\cite{AFL} with \textsc{Kelinci}~\cite{kersten2017poster} interfacing \textsc{Java} bytecodes.
In order to measure the execution cost, the \textsc{Java} programs are instrumented using the ASM bytecode manipulation framework~\cite{bruneton2002asm}.
As a cost metric, we count the \textsc{Java} bytecode instructions, as this metric is substantially used in the related work \cite{antonopoulos2017decomposition, DBLP:conf/ccs/ChenFD17, DBLP:conf/icse/nilizadeh, pasareanu2016multi}. This enables a fair comparison and is often sufficient for our analysis of side channels. Generally, the cost metric in \toolname is easily exchangeable: it supports the execution time, memory consumption, and any other user-specified cost metrics.

\diffuzz stores a cost difference, called \textit{highscore}, internally in the fuzzer and updates it as soon as a mutated input produced a higher cost value.
This highscore is calculated as the cost difference between two executions under the same public input with different secret inputs.
\diffuzz therefore inherently searches for two partitions and tries to maximize cost differences ($\delta$) between them.

In contrast, \toolname searches for arbitrary many secret inputs (bound by $K$) with the same public input.
While \diffuzz assesses the inputs by comparing the cost values of two executions, \toolname assesses the inputs by comparing the cost values of $K$ executions and partitions the costs into classes of \textit{distinguishable} observations.
Consequently, \toolname considers the number of classes (partitions) as a highscore.
Additionally, \toolname maximizes the cost differences ($\delta$) between these partitions as an additional highscore parameter using the minimum distance among the identified partitions.



\begin{figure*}[!htb]
    \centering
    \begin{minipage}{0.3\textwidth}
        \centering
   		\includegraphics[width=1\textwidth]{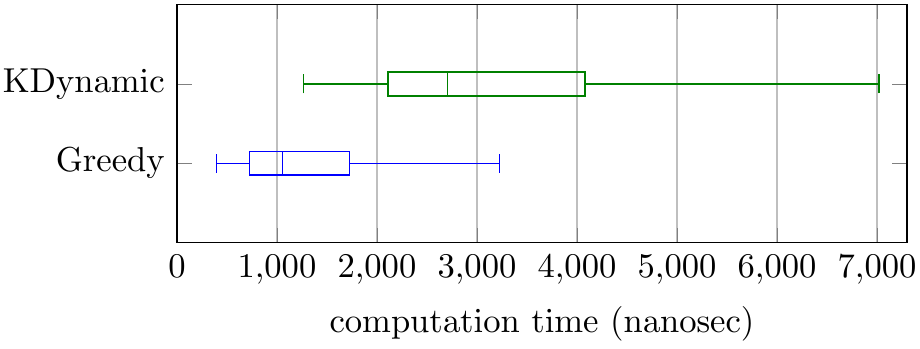}
    \end{minipage}%
    \begin{minipage}{0.35\textwidth}
        \centering
        \includegraphics[width=0.90\textwidth]{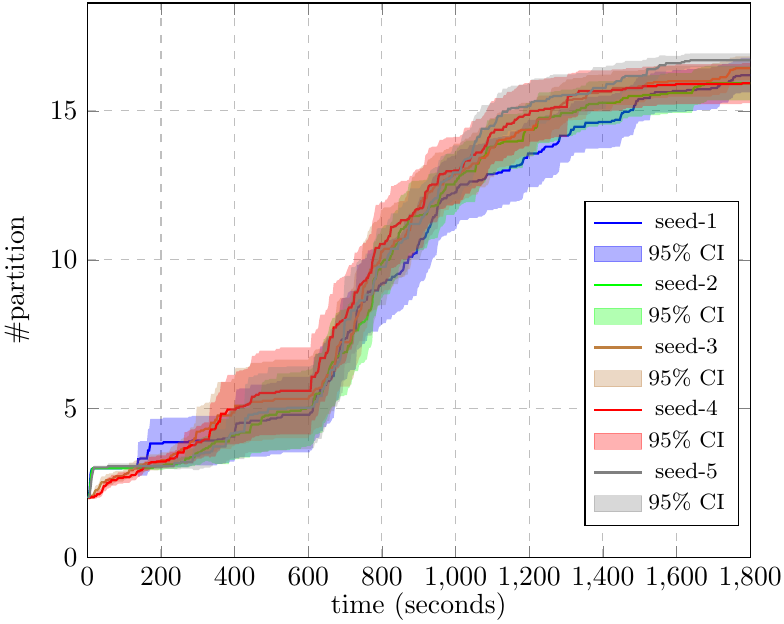}
    \end{minipage}%
    \begin{minipage}{0.35\textwidth}
    	\centering
        \includegraphics[width=0.90\textwidth]{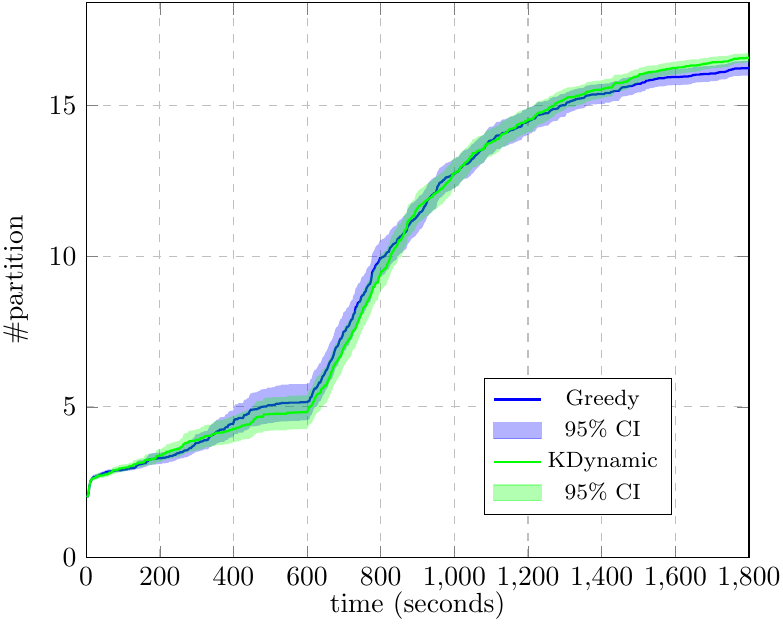}
    \end{minipage}
    \caption{
    (left) Computational complexity of \textit{Greedy} vs. \textit{KDynamic} in isolation;
    (middle) \emph{Eclipse Jetty 1 ($\mathbf{\epsilon=1}$)}: temporal development of 5 different seed inputs with \textit{Greedy};
    (right) \emph{Eclipse Jetty 1 ($\mathbf{\epsilon=1}$)}: temporal development of \textit{Greedy} and \textit{KDynamic} with 5 different seed inputs combined
    (lines and bands show averages and 95\% confidence intervals across 30 repetitions).}
    \label{fig:fuzzing-details}
\end{figure*}

\subsection{Experimental Setup}
We used a virtual machine with Ubuntu $18.04.1\,LTS$
featuring $2x$ Intel(R) Xeon(R) CPU $X5365$ @ $3.00GHz$ with $8GB$ of memory, \textsc{OpenJDK} $1.8.0$\_$191$ and GCC $7.3.0$ to run our experiments.
%
Unless noted otherwise, we execute the experiments with a timeout of
$30$ minutes, a $K$ value of $100$, and an $\epsilon$ value of $1.0$.
Each experiment starts with one (randomly generated) seed input, while we only ensured that this initial input does not crash the application, as required by the underlying fuzzing engine.
We reuse the same seed inputs for comparison with \textit{Greedy} and \textit{KDynamic}, so that both start at the same initial state.
The experiments for RQ1 have been repeated with $5$ different seed inputs to observe the different behavior of our fuzzer for different initial inputs.
For the other experiments (RQ2 and RQ3), we only report the results for one seed input because preliminary experiments as well as the experiments for RQ1 showed that there is no significant variance in the behavior.
Additionally, each experiment has been repeated $30$ times in order to incorporate the randomness of our fuzzing approach.
We averaged the results and calculated the $95$\% confidence intervals as well as the maximum/minimum values.


{\footnotesize
\begin{table*}[ht]
\caption{Comparison of partitioning algorithms (discrepancies are highlighted in \textit{\color{red}red}).}
\centering
\begin{tabu}{|lr|r|l|rr|r|rrr|}
\hline
\multirow{2}{*}{\textbf{Subject}} & \multirow{2}{*}{$\mathbf{\epsilon}$} & \multirow{2}{*}{\textbf{\#Partitions}} & \multirow{2}{*}{\textbf{Algorithm}} & \multirow{2}{*}{$\mathbf{\overline{p}}$} & \multirow{2}{*}{$\mathbf{p_{max}}$} & \multirow{2}{*}{$\mathbf{\delta_{max}}$} & \multicolumn{3}{c|}{\textbf{Time (s)}} \\
& & & & & & & $\overline{p} > 1$ & $p_{max}$ & $t^{min}_{p_{max}}$ \\
\hline


\multirow{2}{*}{Eclipse Jetty 1} & \multirow{2}{*}{1} & \multirow{2}{*}{17} &
Greedy & 16.24 (+/- 0.24) & {17} & 3 & 4.65 (+/- 0.10) & 1436.16 (+/- 53.58) & 675 \\
& & &
KDynamic & 16.59 (+/- 0.15) & {17} & 3 & 4.54 (+/- 0.10) & 1423.65 (+/- 51.89) & 656 \\ \hline

\multirow{2}{*}{Eclipse Jetty 1} & \multirow{2}{*}{4} & \multirow{2}{*}{9} &
Greedy & 8.63 (+/- 0.10) & {9} & 6 & 31.20 (+/- 6.38) & 1428.63 (+/- 56.03) & 697 \\
& & &
KDynamic & 8.65 (+/- 0.12) & {9} & 3 & 31.71 (+/- 6.53) & 1394.75 (+/- 55.87) & 669 \\ \hline


%


%


\multirow{2}{*}{Eclipse Jetty 4} & \multirow{2}{*}{1} & \multirow{2}{*}{9} &
Greedy & 8.51 (+/- 0.09) & 9 & 1 & 3.93 (+/- 0.11) & 1426.89 (+/- 68.86) & 496 \\
& & &
KDynamic & 8.45 (+/- 0.11) & 9 & 1 & 4.15 (+/- 0.11) & 1437.27 (+/- 66.16) & 497 \\ \hline

\multirow{2}{*}{Eclipse Jetty 4} & \multirow{2}{*}{4} & \multirow{2}{*}{4} &
Greedy & 3.94 (+/- 0.04) & \textit{\color{red} 4} & 5 & 28.50 (+/- 4.61) & 1083.39 (+/- 59.33) & 245 \\
& & &
KDynamic & 4.27 (+/- 0.08) & \textit{\color{red} 5} & 2 & 27.21 (+/- 4.07) & 1678.88 (+/- 38.36) & 599 \\ \hline


\multirow{2}{*}{Eclipse Jetty 5} & \multirow{2}{*}{1} & \multirow{2}{*}{1} &
Greedy & 1.00 (+/- 0.00) & 1 & 0 & - & 1.00 (+/- 0.00) & 1 \\
& & &
KDynamic & 1.00 (+/- 0.00) & 1 & 0 & - & 1.00 (+/- 0.00) & 1 \\ \hline

\multirow{2}{*}{Eclipse Jetty 5} & \multirow{2}{*}{4} & \multirow{2}{*}{1} &
Greedy & 1.00 (+/- 0.00) & 1 & 0 & - & 1.00 (+/- 0.00) & 1 \\
& & &
KDynamic & 1.00 (+/- 0.00) & 1 & 0 & - & 1.00 (+/- 0.00) & 1 \\ \hline 


\multirow{2}{*}{Leak Set 1} & \multirow{2}{*}{1} & \multirow{2}{*}{13} &
Greedy & 13.00 (+/- 0.00) & 13 & 92 & 4.20 (+/- 0.07) & 77.53 (+/- 8.31) & 8 \\
& & &
KDynamic & 13.00 (+/- 0.00) & 13 & 92 & 3.77 (+/- 0.09) & 79.77 (+/- 9.36) & 8 \\ \hline

\multirow{2}{*}{Leak Set 2} & \multirow{2}{*}{1} & \multirow{2}{*}{17} &
Greedy & 17.00 (+/- 0.00) & 17 & 92 & 4.13 (+/- 0.07) & 389.46 (+/- 30.03) & 93 \\
& & &
KDynamic & 16.99 (+/- 0.01) & 17 & 92 & 3.73 (+/- 0.12) & 379.14 (+/- 35.43) & 42 \\ \hline

\multirow{2}{*}{Leak Set 3} & \multirow{2}{*}{1} & \multirow{2}{*}{21} &
Greedy & 20.92 (+/- 0.04) & 21 & 92 & 5.01 (+/- 0.01) & 815.82 (+/- 72.93) & 223 \\
& & &
KDynamic & 20.89 (+/- 0.05) & 21 & 92 & 5.00 (+/- 0.00) & 842.37 (+/- 78.11) & 208 \\ \hline

\multirow{2}{*}{Leak Set 4} & \multirow{2}{*}{1} & \multirow{2}{*}{25} &
Greedy & 24.39 (+/- 0.12) & 25 & 92 & 5.00 (+/- 0.00) & 1437.27 (+/- 72.39) & 453 \\
& & &
KDynamic & 24.50 (+/- 0.11) & 25 & 92 & 3.56 (+/- 0.13) & 1389.97 (+/- 71.79) & 268 \\ \hline

\multirow{2}{*}{Leak Set 5} & \multirow{2}{*}{1} & \multirow{2}{*}{29} &
Greedy & 27.71 (+/- 0.17) & 29 & 92 & 4.34 (+/- 0.08) & 1668.63 (+/- 46.38) & 405 \\
& & &
KDynamic & 27.66 (+/- 0.16) & 29 & 92 & 3.77 (+/- 0.11) & 1689.11 (+/- 44.18) & 426 \\ \hline


\multirow{2}{*}{Apache WSS4J} & \multirow{2}{*}{1} & \multirow{2}{*}{17 } &
Greedy & 12.70 (+/- 0.40) & 17 & 3 & 4.93 (+/- 0.13) & 1772.50 (+/- 15.58) & 1079 \\
& & &
KDynamic & 12.72 (+/- 0.41)  & 17 & 3 & 4.89 (+/- 0.13) & 1772.54 (+/- 14.05) & 1301 \\ \hline


\end{tabu}
\label{table:benchmark-cluster}
\end{table*}
}

%

\subsection{Partitioning Algorithms (RQ1)}

In Section \ref{sec:approach}, we propose two partitioning strategies: \textit{KDynamic} and \textit{Greedy}.
The partitioning is a part of our mutant fitness evaluation during fuzzing, which should be as efficient as possible to not slow down the overall fuzzing process.
From a theoretical perspective, \textit{KDynamic} incorporates the maximization of $\delta$, while \textit{Greedy} simply tries to find actual partitions.
Therefore, \textit{KDynamic} is designed to produce high $\delta$ values, while \textit{Greedy} is designed to be fast.
In a preliminary experiment we compared \textit{Greedy} and \textit{KDynamic} in isolation, i.e., just the partitioning for $n=4779$ input files.
We extract the public value and $K=100$ secret values from the input files and then measure the execution time, which each partitioning algorithm needs to calculate the number of partitions.
Figure~\ref{fig:fuzzing-details} (left) shows the statistical comparison: \textit{KDynamic} is (in isolation) significantly slower and the mean execution time is 1.6 times larger.
Please note that the measured time is in nanoseconds, so that the absolute differences between the two execution times might not be essential for the surrounding fuzzing process.

In order to see whether there is a observable difference in the surrounding fuzzing process, we performed additional experiments as described in Section~\ref{sec:subjects} with regard to the number of identified partitions, the cost differences $\delta$, and time to the maximum number of partitions.
Table~\ref{table:benchmark-cluster} shows the results of these experiments.
The column $\#Partitions$ shows the true number of partitions, which should be identified for these subjects.
The columns $\overline{p}$ and $p_{max}$ describe the average number of partitions
with the 95\% confidence interval and
the maximum number of partitions over $30$ runs, respectively.
The column $\delta_{max}$ describes the $\delta$ value for the $p_{max}$.
The column $Time (s): \overline{p} > 1$ shows the average time to identify more than one partition (with the 95\% confidence interval), $Time (s): p_{max}$ shows the average time to the $p_{max}$ value and $t^{min}_{p_{max}}$ shows the minimum time to find the $p_{max}$ over all runs.

Overall, there was no significant difference in the number of partitions identified by both algorithms.
There have been minor discrepancies in favor of the \textit{Greedy} algorithm (as highlighted in red).
Manual inspections showed that \textit{KDynamic} algorithm over-approximates the number of partitions (by one partition) for \textit{Eclipse Jetty 4} ($\epsilon=4$).
The differences in the $\delta$ values are minor or due to different $p_{max}$ values.
For some cases even \textit{Greedy} achieved better $\delta$ values, e.g., \textit{Eclipse Jetty 4} ($\epsilon=4$).
The differences between the computation times of \textit{Greedy} and \textit{KDynamic} are mostly insignificant or due to different $p_{max}$ values.
We therefore conclude that the measured time difference in the preliminary experiments is too small to affect the overall fuzzing results.
The overall fuzzing process with mutations, I/O operations etc. takes longer and outweighs the partitioning effort.


In addition to assessing the final results, we also compared the temporal development of the identified partitions and $\delta$ values between them over five different seed inputs.
Figure \ref{fig:fuzzing-details} (middle+right) shows this exemplary for the subject \textit{Eclipse Jetty 1} with $\epsilon=1$ (the plots for the other subjects can be found in the collected experimental results on our GitHub repository).
We found that there is no significant difference between the $5$ different seed inputs over all subjects, and the two partitioning algorithms perform very similar during the $30$ minutes experiments.

\begin{tcolorbox}[boxrule=1pt,left=1pt,right=1pt,top=1pt,bottom=1pt]
\textbf{Answer RQ1:}
There is no significant difference in terms of the number of partitions or computation time (during fuzzing) between \textit{KDynamic} and \textit{Greedy}.
Interestingly, \textit{Greedy} did produce comparable $\delta$ values.
We choose \textit{Greedy} as the partition algorithm for the our remaining experiments because it provides an under-approximation and is more efficient in isolation.
\end{tcolorbox}

{\footnotesize
\begin{table*}[ht]
\caption{The results of applying \toolname to the \blazer benchmarks (discrepancies are highlighted in \textit{\color{red}red}).}
\centering
\begin{tabu}{|ll|rr|r|rrrr|}
\hline
\multirow{2}{*}{\textbf{Benchmark}} & \multirow{2}{*}{\textbf{Version}} & \multicolumn{2}{c|}{\textbf{\toolname}} & \textbf{\diffuzz} &  \multicolumn{4}{c|}{\textbf{Time (s)}} \\
& & $p_{max}$ & $\delta_{max}$ & $\delta_{max}$ & \toolname, $\overline{p}>1$ & \diffuzz, $\overline{\delta}>0$ & \blazer & \themis \\
\hline
Array & Safe & 1 & 0 & 1 & - & 7.40 (+/- 1.21) & 1.60 & 0.28 \\
Array & Unsafe & 2 & 192 & 195 & 5.70 (+/- 0.21) & 7.40 (+/- 0.93) & 0.16 & 0.23 \\

\rowfont{\em\color{red}}
LoopAndbranch & Safe & 2 & 4 & 4,278,268,702 & 1045.33 (+/- 43.51) & 18.60 (+/- 6.40) & 0.23 & 0.33\\
LoopAndbranch & Unsafe & 2 & 4 & 4,294,838,782 & 1078.63 (+/- 61.04) & 10.60 (+/- 2.62) & 0.65 & 0.16\\

Sanity & Safe & 1 & 0 & 0 & - & - & 0.63 & 0.41  \\
Sanity & Unsafe & 2 & 3,537,954,539 & 4,290,510,883 & 1414.13 (+/- 102.27) & 163 (+/- 40.63)  & 0.30 & 0.17 \\\hline

Straightline & Safe & 1 & 0 & 0 & - & - & 0.21 & 0.49\\
Straightline & Unsafe & 2 & 8 & 8 & 7.47 (+/- 0.18) & 14.60 (+/- 6.53) & 22.20 & 5.30\\

unixlogin & Safe & - & - & 3 & - & 510 (+/- 91.18) & 0.86 & -  \\
unixlogin & Unsafe & 2 & 6,400,000,008 & 3,200,000,008 & 1784.47 (+/- 21.27) & 464.20 (+/- 64.61) & 0.77 & - \\
\hline

modPow1 & Safe & 1 & 0 & 0 & - & - & 1.47 & 0.61\\
modPow1 & Unsafe & 22 & 117 & 3,068 & 4.73 (+/- 0.16) & 4.80 (+/- 1.11) & 218.54 & 14.16\\

modPow2 & Safe & 1 & 0 & 9 & - & - & 1.62 & 0.75 \\
modPow2 & Unsafe & 31 & 1 & 5,206 & 294.70 (+/- 104.66) & 23.00 (+/- 3.48) & 7813.68 & 141.36\\

passwordEq & Safe & 1 & 0 & 0.00 & - & - & 2.70  & 1.10 \\
passwordEq & Unsafe & 93 & 2 & 127 & 4.57 (+/- 0.22) & 8.60 (+/-2.11) &  1.30 & 0.39\\

\hline

k96 & Safe & 1 & 0 & 0  &  - & - & 0.70 & 0.61\\
k96 & Unsafe & 93 & 2 & 3,087,339 & 4.57 (+/- 0.22) & 3.40 (+/- 0.98) & 1.29 & 0.54\\

\rowfont{\em\color{red}}
gpt14 & Safe & 12 & 1 & 517 & 5.00 (+/- 0.00) & 4.20 (+/- 0.80) & 1.43 & 0.46\\
gpt14 & Unsafe & 92 & 2 & 12,965,890 & 5.87 (+/- 0.12) & 4.40 (+/- 1.03) & 219.30 & 1.25\\

login & Safe & 1  & 0 & 0 & - & - & 1.77 & 0.54 \\
login & Unsafe & 17 & 2 & 62 & 7.77 (+/- 0.69) & 10.00 (+/- 2.92) & 1.79 & 0.70 \\
\hline
\end{tabu}
\label{table:benchmark-blazer}
\end{table*}
}

{\footnotesize
\begin{table*}[ht]
\caption{The results of applying \toolname to the \themis benchmarks (discrepancies are highlighted in \textit{\color{red}red}).}
\centering
\begin{tabu}{|ll|rrr|rr|ccr|}
\hline
\multirow{2}{*}{\textbf{Benchmark}} & \multirow{2}{*}{\textbf{Version}} & \multicolumn{3}{c|}{\textbf{\toolname}} & \multicolumn{2}{c|}{\textbf{\diffuzz}} & \multicolumn{3}{c|}{\textbf{\themis}}\\
& & $p_{max}$ & $\delta_{max}$ & Time (s): $\overline{p}>1$ & $\delta_{max}$ & Time (s): $\overline{\delta}>0$ & $\epsilon=64$ & $\epsilon=0$ & Time (s) \\
\hline

Spring-Security & Safe & 1 & 0 & - & 1 & 9.00 (+/- 1.26) & \cmark & \cmark & 1.70 \\
Spring-Security & Unsafe & 2 & 149 & 13.07 (+/- 0.91) & 149 & 8.80 (+/- 1.16) &\cmark & \cmark & 1.09 \\

JDK7-MsgDigest & Safe & 1 & 0 & - & 1 & 15.80 (+/- 3.93) &  \cmark & \cmark & 1.27 \\
JDK6-MsgDigest & Unsafe & 2 & 239 & 4.97 (+/- 0.24) & 34,479 & 7.40 (+/- 1.29) & \cmark & \cmark & 1.33 \\

Picketbox & Safe & 1 & 0 & - & 1 & 29.20 (+/-5.00) &\cmark & \xmark & 1.79 \\
Picketbox & Unsafe & 16 & 2 & 4.20 (+/- 0.14) & 8,794 & 16.80 (+/- 2.58) &\cmark & \cmark & 1.55\\

Tomcat & Safe & 3 & 2 & 1.93 (+/- 0.13) & 14 & 13.80 (+/- 1.29) & \cmark & \xmark & 9.93 \\
\rowfont{\em\color{red}}
Tomcat & Unsafe & 3 & 2 & 1.97 (+/- 0.11) & 37 & 128.60 (+/- 87.20) & \cmark & \cmark & 8.64  \\

\rowfont{\em\color{red}}
Jetty & Safe & 31 & 1 & 4.23 (+/- 0.15) & 8898 & 9.40 (+/- 1.86) & \cmark & \cmark & 2.50 \\
Jetty & Unsafe & 17 & 2 & 4.27 (+/- 0.21) & 16020 & 7.00 (+/- 1.05) & \cmark & \cmark & 2.07\\

orientdb & Safe & 1 & 0 & - & 6 & 3.20 (+/- 0.97) & \cmark & \xmark & 37.99\\
orientdb & Unsafe & 17 & 2 & 1.00 (+/- 0.00) & 19,300 & 3.00 (+/- 0.84) & \cmark & \cmark & 38.09 \\

pac4j & Safe & 2 & 10 & 2.00 (+/- 0.00) & 10 & 5.00 (+/- 1.22) & \cmark & \xmark & 3.97\\
\rowfont{\em\color{red}}
pac4j & Unsafe & 2 & 11 & 2.00 (+/- 0.00) & 11 & 8.00 (+/- 2.76) & \cmark & \cmark &1.85 \\
\rowfont{\em\color{red}}
pac4j & Unsafe* & 2 & 39 & 2.03 (+/- 0.06) & 39 & 10.80 (+/- 5.80) & - & - & - \\

boot-auth & Safe & 2 & 5 & 0.90 (+/- 0.11) & 5 & 5.20 (+/- 0.20) & \cmark & \xmark & 9.12\\
boot-auth & Unsafe & 33 & 3 & 0.93 (+/- 0.09) & 101 & 5.20 (+/- 0.20)  & \cmark & \cmark & 8.31 \\

tourPlanner  & Safe & 1 & 0 & - & 0 & - & \cmark & \cmark & 22.22\\
tourPlanner  & Unsafe & 51 & 1 & 19.97 (+/- 0.24) & 576 & 19.20 (+/- 0.80) & \cmark & \cmark  & 22.01\\

DynaTable & Unsafe & 18 & 2 & 5.83 (+/- 0.13) & 97 & 3.60 (+/- 1.21) & \cmark & \cmark & 1.165 \\

Advanced\_table & Unsafe & 2 & 93 & 11.13 (+/- 1.37) & 97 & 11.20 (+/- 1.62) & \cmark & \cmark & 2.01 \\

OpenMRS & Unsafe & 2 & 206 & 629.43 (+/- 10.41) & 206 & 11.60 (+/- 3.22) & \cmark & \cmark & 9.71 \\

\rowfont{\em\color{red}}
OACC & Unsafe & 18 & 2 & 0.97 (+/- 0.06) & 47 & 7.00 (+/- 1.30) & \cmark & \cmark & 1.83 \\

\hline
\end{tabu}
\label{table:benchmark-themis}
\end{table*}
}

{\footnotesize
\begin{table*}[ht]
\caption{Results on \diffuzz's additional examples.}
\centering
\begin{tabu}{|ll|rrr|rr|}
\hline
\multirow{2}{*}{\textbf{Benchmark}} & \multirow{2}{*}{\textbf{Version}} & \multicolumn{3}{c|}{\textbf{\toolname}} & \multicolumn{2}{c|}{\textbf{\diffuzz}} \\
& & $p_{max}$ & $\delta_{max}$ & Time (s): $\overline{p}>1$ & $\delta_{max}$ & Time (s): $\overline{\delta}>0$ \\ 
\hline
CRIME & unsafe & 33 & 1 & 5.00 (+/- 0.00) & 782 & 7.40 (+/- 1.12) \\
ibasys (imageMacher) & unsafe & 9 & 9 & 48.30 (+/- 8.21) & 262 & 6.20 (+/- 0.66)\\
\hline

Apache ftpserver Clear & safe & 1 & 0 & - & 1 & 7.20 (+/- 1.24) \\
Apache ftpserver Clear & unsafe & 17 & 2 & 0.87 (+/- 0.12) & 47 & 6.80 (+/- 1.07) \\

Apache ftpserver MD5 & safe & 1 & 0 & - & 1 & 4.20 (+/- 1.93) \\
Apache ftpserver MD5 & unsafe & 5 & 9 & 0.90 (+/- 0.11) & 151 & 2.80 (+/- 1.11)  \\

Apache ftpserver SaltedPW & (safe) & 43 & 1 & 12.10 (+/- 0.19) & 198 & 2.20 (+/- 0.73) \\
Apache ftpserver SaltedPW & unsafe & 42 & 1 & 11.93 (+/- 0.18) & 193 & 3.60 (+/- 1.08)\\
Apache ftpserver SaltedPW* & unsafe & 54 & 1 & 11.90 (+/- 0.17) & 178 & 5.40 (+/- 0.98) \\

Apache ftpserver StringUtils & safe & 1 & 0 & - & 0 &  - \\ 
Apache ftpserver StringUtils & unsafe & 17 & 3 & 4.27 (+/- 0.16) & 53 & 3.00 (+/- 1.05) \\

AuthMeReloaded & safe & 1 & 0 & - & 1 & 7.60 (+/- 0.75) \\
AuthMeReloaded & unsafe & 5 & 3 & 2.00 (+/- 0.00) & 383 & 9.20 (+/- 1.96) \\
\hline
\end{tabu}
\label{table:benchmark-diffuzz}
\end{table*}
}
{\footnotesize
\begin{table*}[ht]
\caption{The results of applying \toolname to the RSA subjects in \textsc{MaxLeak}~\cite{pasareanu2016multi} (\textbf{\color{red}red} highlighted exceeded the budget: timeout of 1 hour or memory of 8GB, \textit{\color{blue}blue} highlighted partitions are below the maximum possible observation).}
\centering
\begin{tabu}{|rr|r|rrrr|rr|rr|}
\hline
\multirow{2}{*}{\textbf{Modulo}} & \multirow{2}{*}{\textbf{Len}} & \multirow{2}{*}{\textbf{\#Partitions}} & \multicolumn{4}{c|}{\textbf{\toolname ($\mathbf{\epsilon}$=0, 1h)}} & \multicolumn{2}{c|}{\textbf{MaxLeak (default)}} & \multicolumn{2}{c|}{\textbf{MaxLeak (No solver)}} \\
& & & $\overline{p}$ & $p_{max}$ & Time (s): $p_{max}$ & $t_{min}$ & \#Obs & Time (s) & \#Obs & Time (s) \\
\hline


1717 & 3 & 7 & 7.00 (+/- 0.00) & 7 & 1.00 (+/- 0.00) & 1 & 6 & 20.892 & 9 & 1.047 \\
1717 & 4 & 10 & 10.00 (+/- 0.00) & 10 & 7.43 (+/- 0.45) & 5 & 9 & 152.332 & 12 & 1.370 \\
1717 & 5 & 13 & 13.00 (+/- 0.00) & 13 & 20.40 (+/- 3.87) & 6 & 12 & 839.788 & 15 & 2.916 \\
1717 & 6 & 16 & 16.00 (+/- 0.00) & 16 & 294.60 (+/- 53.17) & 22 & 15 & \textbf{\color{red}3731.328} & 18 & 8.006 \\
1717 & 7 & 19 & 18.37 (+/- 0.25) & 19 & 2484.30 (+/- 451.42) & 385 & \multicolumn{2}{c|}{\textbf{\color{red}> 4 h}} & 21 & 19.241 \\

1717 & 8 & 22 & 20.43 (+/- 0.45) & 22 & 3168.07 (+/- 303.47) & 508 & \multicolumn{2}{c|}{\textbf{\color{red}> 4 h}} & 24 & 91.821 \\
1717 & 9 & 25 & 22.20 (+/- 0.36) & \textit{\color{blue}24} & 3489.03 (+/- 169.19) & 1009 & \multicolumn{2}{c|}{\textbf{\color{red}> 4 h}} & \multicolumn{2}{c|}{\textbf{\color{red}> 8 GB}} \\
1717 & 10 & 28 & 24.40 (+/-  0.49) & \textit{\color{blue}27} & 3548.63 (+/- 57.73) & 2929 & \multicolumn{2}{c|}{\textbf{\color{red}> 4 h}} & \multicolumn{2}{c|}{\textbf{\color{red}> 8 GB}} \\
\hline
834443 & 3 & 7 & 7.00 (+/- 0.00) & 7 & 13.40 (+/- 1.96) & 8 & 6 & 7.416 & 9 & 1.188 \\
834443 & 4 & 10 & 10.00 (+/- 0.00) & 10 & 40.33 (+/- 12.14) & 6 & 9 & 42.684 & 12 & 1.385 \\
834443 & 5 & 13 & 12.93 (+/- 0.09) & 13 & 645.70 (+/- 329.43) & 74 & 12 & 215.929 & 15 & 2.953 \\
834443 & 6 & 16 & 15.40 (+/- 0.20) & 16 & 2711.87 (+/- 433.23) & 271 & 15 & 936.921 & 18 & 7.511 \\
834443 & 7 & 19 & 16.80 (+/- 0.33) & 18 & 3227.60 (+/- 275.29) & 952 & 18 & \textbf{\color{red}4021.150} & 21 & 19.068 \\
834443 & 8 & 22 & 17.93 (+/- 0.54) & 22 & 3556.70 (+/- 83.44) & 2301 & \multicolumn{2}{c|}{\textbf{\color{red}> 4 h}} & 24 & 96.360 \\
834443 & 9 & 25 & 20.13 (+/- 0.59) & \textit{\color{blue}24} & 3572.83 (+/- 37.16) & 3110 & \multicolumn{2}{c|}{\textbf{\color{red}> 4 h}} & \multicolumn{2}{c|}{\textbf{\color{red}> 8 GB}} \\
834443 & 10 & 28 & 21.83 (+/- 0.46) & \textit{\color{blue}24} & 3504.13 (+/- 121.70) & 1845 & \multicolumn{2}{c|}{\textbf{\color{red}> 4 h}} & \multicolumn{2}{c|}{\textbf{\color{red}> 8 GB}} \\
\hline
1964903306 & 3 & 7 & 6.47 (+/- 0.18) & 7 & 2228.30 (+/- 542.13) & 119 & 6 & 12.167 & 9 & 1.085 \\
1964903306 & 4 & 10 & 8.67 (+/- 0.19) & 10 & 3494.30 (+/- 203.69) & 429 & 9 & 70.805 & 12 & 1.535 \\
1964903306 & 5 & 13 & 10.70 (+/- 0.19) & 12 & 3594.00 (+/- 11.56) & 3420 & 12 & 2306.261 & 15 & 3.391 \\
1964903306 & 6 & 16 & 12.90 (+/- 0.11) & \textit{\color{blue}13} & 1337.90 (+/- 443.89) & 206 & \multicolumn{2}{c|}{\textbf{\color{red}> 4 h}} & 18 & 7.506 \\
1964903306 & 7 & 19 & 14.10 (+/- 0.27) & \textit{\color{blue}15} & 2984.67 (+/- 362.05) & 503 & \multicolumn{2}{c|}{\textbf{\color{red}> 4 h}} & 21 & 19.486 \\
1964903306 & 8 & 22 & 15.33 (+/- 0.36) & \textit{\color{blue}17} & 3398.37 (+/- 204.45) & 1411 & \multicolumn{2}{c|}{\textbf{\color{red}> 4 h}} & 24 & 98.325 \\
1964903306 & 9 & 25 & 16.30 (+/- 0.51) & \textit{\color{blue}19} & 3562.33 (+/- 54.24) & 2819 & \multicolumn{2}{c|}{\textbf{\color{red}> 4 h}} & \multicolumn{2}{c|}{\textbf{\color{red}> 8 GB}} \\
1964903306 & 10 & 28 & 17.30 (+/- 0.48) & \textit{\color{blue}20} & 3559.67 (+/- 77.72) & 2390 & \multicolumn{2}{c|}{\textbf{\color{red}> 4 h}} & \multicolumn{2}{c|}{\textbf{\color{red}> 8 GB}} \\

\hline
\end{tabu}
\label{table:rsa-comparison}
\end{table*}
}

\subsection{Comparison with \blazer, \themis, and \diffuzz (RQ2)}

\toolname is useful for both the detection and quantification of side channels (SC),
and hence, we can compare \toolname to detection techniques.
We compare \toolname with \blazer~\cite{antonopoulos2017decomposition}, \themis~\cite{DBLP:conf/ccs/ChenFD17}, and \diffuzz~\cite{DBLP:conf/icse/nilizadeh}, the three state-of-the-art SC detectors.
Furthermore, \toolname's implementation is an extension of \diffuzz, and therefore can serve as a baseline with regard to side-channel detection.

The benchmark of the related studies include subjects mostly in two variants: \textit{safe} and \textit{unsafe}.
A \textit{safe} variant is supposed to not show any side-channel vulnerability, while the \textit{unsafe} variant is known to include a vulnerability.
For \textit{unsafe} subjects,
\toolname should identify at least $2$ partitions.
This indicates some measurable differences in the public observations depend
on the secret values.
Consequently, for \textit{safe} subjects, \toolname should identify exactly
one partition.

Tables~\ref{table:benchmark-blazer}, \ref{table:benchmark-themis}, and \ref{table:benchmark-diffuzz} show the corresponding results.
The columns for \toolname only show the $p_{max}$ and $\delta_{max}$
since these parameters are the most relevant ones in detecting side channels.
In order to compare the analysis time, we report the time until \toolname identifies at least two partitions, which compares well with \diffuzz's time parameter $\delta\!>\!0$.
Our default value of $K\!=\!100$ was not applicable to a few subjects: \textit{LoopAndBranch}, \textit{Sanity}, and \textit{UnixLogin} because they represent relatively expensive executions.
As mentioned in Section \ref{sec:approach}, the fuzzer performs $K$ concrete executions to collect the observations. 
Since a large value of $K$ strongly influences the time for input assessments, we ultimately used $K\!=\!20$ for \textit{LoopAndBranch}, $K\!=\!10$ for \textit{Sanity}, and $K\!=\!2$ for \textit{UnixLogin}, which are sufficient for
detecting side channels.

\subsubsection{Results with regard to Partitions (RQ2/part1).}
In all unsafe subjects, \toolname identifies at least $2$ partitions, for which \diffuzz also identifies a high $\delta$ value.
There are discrepancies with \blazer and \themis classifications/results (see red highlights in Table \ref{table:benchmark-blazer} and \ref{table:benchmark-themis}), but these have been already reported with \diffuzz in \cite{DBLP:conf/icse/nilizadeh} and essentially represent false classifications by \blazer and \themis.

\toolname's capabilities go beyond identifying side-channel vulnerabilities.
For example, \blazer's \textit{login} subjects (see Table~\ref{table:benchmark-blazer}) handle the checking of a secret password (in our case with a fixed length of $16$ characters).
Our technique identifies $17$ partitions with a minimum cost difference of two bytecodes between partitions ($\delta=2$).
This represents the partitions ranging from the $0$ (no) prefix character matching to all $16$ prefix characters matching for a given public input.
Therefore, \toolname gives stronger evidence that this vulnerability is
actually exploitable.
Such a statement cannot be inferred with the results from \diffuzz because it only provides the maximum $\delta$ value between any two partitions.

Another example is the \textit{unsafe} variant of \textit{Spring-Security} in the \themis benchmark, which had been already mentioned in Section \ref{sec:overview}.
\diffuzz, \themis, and \toolname conclude that it is unsafe.
However, the provided information to assess the severity of this subject is quite different.
In this subject, the secret password also has a fixed length of $16$ characters.
The String comparison is very similar to the safe variant of \textit{Eclipse Jetty} (see Figure \ref{fig:jetty-streql}).
The fuzzing driver ensures that public and secrets Strings have the same length, so that the comparison is based on the content of the Strings.
Hence, one would expected that no cost differences are possible.
But, there is an additional length-check in the beginning: the \textit{expected} lengths of both
Strings are compared based on the \texttt{String.getBytes("UTF-8")} function.
There is an early return, if the lengths of the resulting byte arrays do not match.

While \diffuzz reports a cost difference of $149$ bytecodes, \toolname reports that there are only  $2$ partitions that are $149$ bytcodes far from each other.
Therefore, \diffuzz indicates some vulnerability, whereas \toolname shows that the strength of the leak is weak, and the vulnerability is unlikely to leak the complete secret.
A closer look into the identified partitions reveals that the side channels only leak whether the secret String contains a special character or not.
Although the quantification of information leaks by \toolname is an under-approximation of the true number of partitions (because of its dynamic nature), it significantly supports the understanding of the vulnerability and
the strength of leaks.

\begin{tcolorbox}[boxrule=1pt,left=1pt,right=1pt,top=1pt,bottom=1pt]
\textbf{Answer RQ2 (Part 1/2):}
\toolname detects the same vulnerabilities similar to state-of-the-art techniques.
Furthermore, \toolname provides additional information about the strength
of leaks and the exploitability of vulnerabilities.
\end{tcolorbox}

\subsubsection{Results with regard to Analysis Time (RQ2/part2).}
Comparing the fuzzing time to the first inputs that reveal more than $1$ partition,
\toolname is considerably slower than the other techniques in some cases.
The large $K$ value in our experiments (usually $K=100$) triggers a large number of concrete program executions during input assessment in fuzzing.
If the program executions are expensive as well, then this can slow down the overall fuzzing campaign.
On the one hand, a large value for $K$ enables \toolname to identify up to $K$ partitions and may lead to a faster exploration via considering multiple secret values.
On the other hand, the large value slows down the overall fuzzing process,
as the input assessments take longer.
The choice of an appropriate value for $K$ remains a trade-off between many partition explorations and a few partition exploitations.

We also observed that in some cases \toolname is significantly faster than the static analysis techniques \blazer and \themis (e.g., \textit{Straightline unsafe} and \textit{modPow1/2 unsafe}).
As reported in \blazer~\cite{antonopoulos2017decomposition}, for the long-running benchmarks \blazer suffers from the combinatorial growth of necessary expression comparisons.
\themis can improve but still suffers for complex benchmarks.
Note that both techniques use taint analysis that is known to be computationally expensive for languages with dynamic features such as \textsc{Java}~\cite{landman2017challenges}.
\toolname (as well as \diffuzz) uses a dynamic analysis, which outperforms static analysis in such cases.

\begin{tcolorbox}[boxrule=1pt,left=1pt,right=1pt,top=1pt,bottom=1pt]
\textbf{Answer RQ2 (Part 2/2):}
Large values for $K$ may slow down \toolname, but eventually, enable the exploration of many partitions.
\toolname outperforms static analysis on complex benchmarks.
\end{tcolorbox}

\subsection{Comparison to \textsc{MaxLeak}~\cite{pasareanu2016multi} on RSA subjects (RQ3)}
\label{subsec:results-rsa}

Pasareanu et al.~\cite{pasareanu2016multi} quantify information leaks using symbolic execution and model counting (MaxSMT).
In particular, they evaluate their approach on the implementations of fast modular exponentiation.
Their cost model does not count executed bytecode instructions, but counts the number of visited branches.
To match with their evaluations, we customized our cost model, reduced $\epsilon$ to zero, and increased the timeout of the experiments to one hour. To enable a fair comparison, we reproduced \textsc{MaxLeak}'s results on our experiment setup.

Table~\ref{table:rsa-comparison} shows the results for these experiments (similar to Figure 9 in \cite{pasareanu2016multi}).
The column \textit{Modulo} denotes the modulo value used for the modulo exponentiation, while the column \textit{Len} denotes the bitvector length of the secret.
The column \textit{\#Partitions} shows the groundtruth for the number of identifiable partitions.
The true number of partitions is formulated to be $3$*(\textit{Len}-$1$) in~\cite{pasareanu2016multi} for every experiment.
We noticed that the formulation is for \textit{Len}${>}1$, while
we also consider \textit{Len}${=}1$, which leads us to find one more
partition.
Therefore, we report the \textit{\#Partitions} as: $3$*(\textit{Len}-$1$)+1.
For \toolname, we report
the average number of identified partitions $\overline{p}$ with the 95\% confidence intervals,
the maximum number of partitions $p_{max}$,
the average time to $p_{max}$, and
the minimum time to $p_{max}$ over all $30$ runs.
The results by~\cite{pasareanu2016multi} come in two modes:
\textit{default} and \textit{no solver}.
The second one represents symbolic execution without filtering infeasible path constraints, which is faster but over-approxi\-mates the number of observations.
For each mode, the columns show the number of \textit{different}
observations (i.e., number of partitions) and the combined time for
symbolic execution and MaxSMT solving.

The results in Table~\ref{table:rsa-comparison} show that for modulo=$1717$, \toolname reliably identifies the correct number of partitions up to a bitvector length of $8$.
For the reported longer secrets, \toolname identifies a close under-approximation within the given timeout of 1 hour.
\textsc{MaxLeak} (default) shows longer runtimes and already exceeds the timeout for the bitvector length of $6$.
Note that we manually stopped the executions after 4 hours (3 hours later than the timeout) so that we cannot report the exact times for longer running subjects.
\textsc{MaxLeak} (no solver) is comparably very fast, but over-ap\-proxi\-mates the right number of partitions.
It eventually exceeds our memory budget of 8 GB for bitvector lengths higher or equal to $9$.
With higher values for \textit{Modulo}, we can still confirm that \toolname produces the best results.
However, it also slows down so that, e.g., for $Modulo=1964903306$, \toolname can only produce reliable estimations for bitvectors lengths up to $5$.
For larger lengths, \toolname still can provide good lower-bounds for the true number of partitions.

\begin{tcolorbox}[boxrule=1pt,left=1pt,right=1pt,top=1pt,bottom=1pt]
\textbf{Answer RQ3: }
Due to its dynamic analysis, \toolname is more scalable than \textsc{MaxLeak}.
We also find that \toolname with lower-bound guarantees performs well on the RSA subjects and has precision comparable to \textsc{MaxLeak}~\cite{pasareanu2016multi} that uses symbolic execution with model counting.
\end{tcolorbox}

\section{Related Work}
\label{sec:related}
\subsubsection*{Fuzzing and Testing for Side-Channel Detection}
Various testing techniques~\cite{DBLP:conf/icse/nilizadeh,milushev2012noninterference,DBLP:journals/corr/abs-1904-07280}
have been employed to identify side channels. We compared our approach
to \diffuzz thoroughly in Section~\ref{sec:experiment}.
Milushev et al.~\cite{milushev2012noninterference} adapt symbolic executions
with self-composition techniques~\cite{barthe2004secure} to detect the noninterference violations due to direct and indirect information leaks.
Recently, He al.~\cite{DBLP:journals/corr/abs-1904-07280} adapt greybox
fuzzing techniques with self-composition specifications to detect
side-channel leaks. Both of these works consider the noninterference
notion of security, while our approach also quantifies the amount of information leaks.

\subsubsection*{Quantitative Analysis of Side Channels}
Quantitative information
flow~\cite{backes2009automatic,smith2009foundations,KB07,DBLP:conf/cav/Tizpaz-NiariC019} has been used for
measuring the strength of side channels and mitigating potential leaks.
We compared our approach with Pasareanu et al.~\cite{pasareanu2016multi}
in Section~\ref{sec:experiment}.
Smith~\cite{smith2009foundations} studies various quantitative notions
of confidentiality and finds that they have limitations in evaluating
the resulting threats if an attacker can try one ideal guess. Therefore, Smith~\cite{smith2009foundations}
introduces min entropy and shows its usefulness in quantifying
the strength of leaks. Our approach adapts greybox fuzzing with
partitioning algorithms to characterize a lower-bound on min entropy.
Backes et al.~\cite{backes2009automatic}
present an approach based on finding the equivalence relation over secret
inputs. They cast the problem of finding the equivalence relation as a
reachability problem and use model counting to quantify information leaks.
Their approach considers only direct information leaks and works for
a small program, limited to a few lines of code.

\subsubsection*{Dynamic Analysis for Side Channels}
Dynamic analysis has been used for analyzing side
channels~\cite{FuncSideChan20,DBLP:conf/icse/nilizadeh,10.1145/3395363.3397365}.
Tizpaz-Niari et al.~\cite{FuncSideChan20} developed a data-driven
approach to debug timing side channels. They define functional
side channels where an attacker may observe the response times
of application on many public inputs (potentially unbounded).
While their approach
uses machine learning-based clustering to detect side channels and localize
root causes, their clustering is after input generations.
Thus, their approach does not guarantee to maximize the number of clusters.
In addition, their approach is limited to a setting
where an attacker passively observes the response times,
whereas we consider a more realistic setting where an attacker
can pick public inputs.

\subsubsection*{Static Analysis for Side Channels}
Static analysis techniques are adapted in various works
to detect side channels~\cite{DBLP:conf/ccs/ChenFD17,antonopoulos2017decomposition,
doychev2015cacheaudit,wang2017cached}.
\themis~\cite{DBLP:conf/ccs/ChenFD17} uses Cartesian Hoare Logic~\cite{barthe2004secure}
with taint analysis to detect side channels.
Similar to \diffuzz, we show that our approach outperforms
\themis, and it also characterizes the amounts of leaks.

\subsubsection*{Adaptive Side Channels}
Adaptive side channels allow attackers to use previous observations to pick
an ideal public input in each step of an attack~\cite{phan2017synthesis,bang2016string}.
Phan et al.~\cite{phan2017synthesis} consider synthesizing
adaptive side channels where in each step,
the attacker chooses the best
public input that maximizes the amount of leaks.
They reduce the problem of finding $k$ ideal public inputs for compromising
secrets to an optimization problem using symbolic
execution and MaxSMT~\cite{10.1007/11814948_18}.
In contrast, we focus on threat models
with one ideal public input and adapt greybox fuzzing
to characterize min entropy. Thus, Adaptive attacks are beyond the scope
of our threat model.
Despite it, we show that QFuzz can indicate the possibility of adaptive attacks.
We left further analysis as future work.

\subsubsection*{Side Channels from Micro-Architecture}
Recently, significant research has been conducted to address transient
execution vulnerabilities such as Spectre attacks~\cite{Kocher2018spectre,Guarnieri2020spectector}. These attacks exploit
timing side channels due to speculative executions from micro-architecture
sources and can compromise confidentiality even if the source code is
completely secure.
Our approach focuses on vulnerabilities in the source code since
a secure software in the source code level is a precondition to have secure systems.
Since the total elimination of Spectre vulnerabilities is not practical
(due to performance concerns), our quantitative approach
can potentially be useful to guarantee security for micro-architecture
designs.

\subsubsection*{Side Channels from Compiler Optimization}
Brennan et al.~\cite{BrennanICSE20} show
potential timing leaks induced from \textsc{JIT} optimizations via
actual runtime observations, which is different than our analysis based
on vulnerabilities in the source code. We opted to count the bytecode,
instead of actual execution times, since the metric is substantially
used in the related work for a fair comparison
and is often sufficient for our analysis of side channels in code. We note that
the cost metric in QFuzz is easily exchangeable: it supports the execution time,
bytecode abstraction, memory consumption, and any other user-specified cost metrics.

\section{Discussion}
\label{sec:limit}

\subsubsection*{Limitations}
Our approach requires a driver program, an upper-bound on partitions $K$, and the tolerance parameter $\epsilon$ as inputs.
In our evaluation, we discuss the trade-off to pick appropriate values for $K$.
For the tolerance, we usually set $\epsilon=1$ to be conservative and comparable to existing works.
In practice, developers can run experiments with different $\epsilon$ values to adapt for the specific context.

A dynamic analysis, as used in this work, often scales well for large applications and can handle dynamic features.
However, such an analysis can lead to false negatives.
Nonetheless, \toolname provides precise under-approximations, and
our evaluation shows that \toolname provides a good indication of resulting threats.
%

The leveraged min entropy assumes that an attacker can try one time, and it may fail to evaluate the resulting threats if an attacker can try multiple times adaptively.
However, we find that a higher number of partitions can serve as warning signs for feasible side-channel attacks over multiple trials such as adaptive attacks.


\subsubsection*{Threats to Validity}
To ensure that our evaluation does not lead to invalid conclusions (i.e., internal validity), we followed the established guidelines \cite{Arcuri2014AssessFuzzing,Klees2018EvaluateFuzzing} to incorporate the randomness in fuzzing.
In particular, we repeated all experiments $30$ times, showed the error margins, performed experiments with multiple seed inputs, and considered not only the final results but also the temporal development.
To illustrate that our results generalize (i.e., external validity), we applied \toolname on a large range of subjects from existing work including micro- and macro-benchmarks.
We specifically used \toolname to analyze real-world programs and also identified a previously unknown vulnerability.
We leveraged evaluation metrics like the number of identified partitions and the time to identify more than one partition because they quantify the information leakage for practical usage and represent how fast our technique can make progress.

\subsubsection*{Usage Vision}
Based on the experiences during our evaluation, we envision the following usage for \toolname.
As a first step, use \toolname to detect side-channel vulnerabilities, and at the same time quantify their strengths.
The obtained quantification facilitates the prioritization of the detected vulnerabilities.
Secondly, focus on the manual investigation to mitigate the side channels guided by the results of the first step.

\section{Conclusion}
\label{sec:conclusion}
We presented \toolname, the first quantitative greybox fuzzing approach for measuring the strength of side-channel leaks.
We have shown that \toolname outperforms existing state-of-the-art techniques for detection and quantification of side-channel vulnerabilities.
Furthermore, \toolname found a zero-day vulnerability in a security-critical \textsc{Java}
library that has since been fixed.
For future work, we aim to extend \toolname to quantitatively evaluate
the resulting threats from an attacker who can try multiple times adaptively.

\begin{tcolorbox}[boxrule=1pt,left=1pt,right=1pt,top=1pt,bottom=1pt]
Our open-source tool \toolname and all experimental subjects are publicly accessible:
\begin{itemize}
	\item \url{https://github.com/yannicnoller/qfuzz}
	\item \url{https://doi.org/10.5281/zenodo.4722965}
\end{itemize}
\end{tcolorbox}

\bibliographystyle{ACM-Reference-Format}
\bibliography{papers}

\end{document}